%% file: main.tex
\documentclass[conference,letterpaper,preprint]{IEEEtran}

\usepackage[dvipsnames]{xcolor}
\usepackage[utf8]{inputenc} 
\usepackage[T1]{fontenc}
\usepackage{url}
\usepackage{ifthen}
\usepackage{cite}
\usepackage{bm}
\usepackage[cmex10]{amsmath} % Use the [cmex10] option to ensure complicance
\usepackage[%pagebackref, 
hidelinks, colorlinks=true, citecolor=black!30!blue,linkcolor=black]{hyperref}
\usepackage{graphicx}

\usepackage{wrapfig}
\usepackage[utf8]{inputenc} % allow utf-8 input
\usepackage[T1]{fontenc}    % use 8-bit T1 fonts
\usepackage{hyperref,xcolor}       % hyperlinks
\usepackage{url}            % simple URL typesetting
\usepackage{booktabs}       % professional-quality tables
\usepackage{amsfonts}       % blackboard math symbols
\usepackage{nicefrac}       % compact symbols for 1/2, etc.
\usepackage{microtype}      % microtypography
\usepackage{xcolor}         % colors
\usepackage{comment}

\usepackage{amsmath,amsfonts,amssymb}
\usepackage{graphicx}

\usepackage{algorithm}
\usepackage{algorithmic}

\usepackage{multirow}

\input{defns}

\begin{document}
\title{
DeCompress: Denoising via Neural Compression
} 

% %%% Single author, or several authors with same affiliation:
% \author{%
%  \IEEEauthorblockN{Author 1 and Author 2}
% \IEEEauthorblockA{Department of Statistics and Data Science\\
%                    University 1\\
 %                   City 1\\
  %                  Email: author1@university1.edu}% }

%%% Several authors with up to three affiliations:
\author{%
  \IEEEauthorblockN{Ali Zafari$^*$, Xi Chen$^*$, Shirin Jalali}
  \IEEEauthorblockA{Department of Electrical and Computer Engineering \\
                    Rutgers University\\
                    Piscataway, NJ, USA\\
                    Email: \{ali.zafari, xi.chen15, shirin.jalali\}@rutgers.edu}
}

\maketitle
\def\thefootnote{*}\footnotetext{Equal contribution.}
%%%%%%
%% Abstract: 
%% If your paper is eligible for the student paper award, please add
%% the comment "THIS PAPER IS ELIGIBLE FOR THE STUDENT PAPER
%% AWARD." as a first line in the abstract. 
%% For the final version of the accepted paper, please do not forget
%% to remove this comment!
%%

\begin{abstract}
Learning-based denoising algorithms achieve state-of-the-art performance across various denoising tasks. However, training such models  relies on access to large training datasets consisting of clean and noisy image pairs. On the other hand, in many imaging applications, such as microscopy, collecting ground truth images is often infeasible. To address this challenge, researchers have recently developed algorithms that can be trained without requiring access  to ground truth data. However, training such models remains  computationally challenging and still requires access to large  noisy training  samples. In this work, inspired by compression-based denoising and recent advances in neural compression, we propose a new compression-based denoising algorithm, which we name DeCompress, that i) does not require access to ground truth images, ii) does not require access to large training dataset - only a single noisy image is sufficient, iii) is robust to overfitting, and iv) achieves superior performance compared with zero-shot or unsupervised learning-based denoisers.
\end{abstract}

\section{Introduction}
 Image denoising studies the problem of recovering image $X^n\in\Xc^n$  from noisy measurements $Y^n\in\mathbb{R}^n$, where
\begin{equation*}
  Y^n = X^n + Z^n; \quad Z^n \sim \Nc(0, \sigma^2I),
\end{equation*}
At the core of all image denoising algorithms lies a method to leverage the inherent structure of the source. Classical denoising algorithms  relied on  handcrafted features identified by domain experts \cite{
% wiener1964extrapolation,
% donoho1994ideal,
mallat1999wavelet,donoho2002noising,portilla2003image,elad2006image,roth2009fields,gu2014weighted}. Recently, the effectiveness of deep learning methods in capturing complex source structure from training data has led researchers to develop various learning-based denoisers \cite{zhang2017beyond,zhang2018ffdnet,lefkimmiatis2018universal,zhang2020residual,liang2021swinir}. Although these methods achieve state-of-the-art performance on a benchmark datasets, they still face limitations, such as reliability concerns. Additionally, learning-based methods typically require extensive training datasets consisting of clean and noisy image pairs. In many practical imaging applications in remote sensing \cite{safonova2023ten}, microscopy \cite{
% xing2017deep,
belthangady2019applications} and astronomical imaging \cite{baron2019machine}, however, not only is the availability of training images   limited, but obtaining  ground truth images is often impossible. 

To address some of these limitations, recent studies have introduced  denoising algorithms  that  can be trained exclusively on noisy images without requiring  access to paired noisy and clean images \cite{soltanayev2018training,lehtinen2018noise2noise,krull2019noise2void,batson2019noise2self,kim2021noise2score}.  However, such methods still require access to large training datasets, and their performance is suboptimal compared to methods that have access to clean images. Furthermore, due to the complexity of the neural net structures required in the absence of clean images, training such models often is challenging. 

The  challenges with existing learning-based methods highlight the  need to   explore alternative approaches. Compression algorithms provide one such alternative direction. The goal of a compression code is to efficiently represent signals by leveraging their underlying structure, similar to denoising algorithms. Consider  lossy compression of a noisy signal $Y^n = X^n + Z^n$ at a distortion level proportional to the variance of the additive noise $Z^n$. In this scenario, the clean signal $X^n$ emerges as a strong candidate for reconstructing $Y^n$ due to its inherent structured nature, enabling efficient representation. This intuitive connection between denoising and compression, which naturally leads to a compression-based denoising approach, has been formally established in the literature \cite{donoho2002kolmogorov,weissman2005empirical}. Despite the  theoretical foundations supporting compression-based denoising, its practical application beyond finite-alphabet sources has remained limited.

Joint optimization of the parameters of an autoencoder while constraining the rate of its bottleneck \cite{balle2020nonlinear}, coined as neural compression, has gained a lot of attention in recent year. It has led to development of learning-based lossy data compression methods \cite{yang2023introduction} for general sources by only having access to their samples. Since then, learning-based neural  compression algorithms have advanced to outperform traditional hand-engineered compression schemes for high dimensional data such as images \cite{balle2017endtoend,balle2018variational,minnen2018joint}, in terms of rate-distortion performance.  The superior rate-distortion performance of neural lossy image compression, coupled with its learning-based nature, motivates the following question:
\begin{center}
\emph{Can we employ neural lossy compression codes to develop efficient image denoising algorithms?}
\end{center}

To optimize the performance of compression-based denoising, it is crucial to adjust the compression code's operating point according to the noise characteristics. In our proposed neural compression-based denoiser, this adjustment is achieved by tuning the Lagrangian coefficient associated with the rate term. Unlike other learning-based methods such as Noise2Noise \cite{lehtinen2018noise2noise}, Noise2Void \cite{krull2019noise2void}, Noise2Self \cite{batson2019noise2self}, and Noise2Score \cite{kim2021noise2score}, which require either multiple noisy versions of the same image or extensive noisy image datasets for training, our proposed neural compression-based denoiser eliminates the need for both. Instead, our method can effectively train on a very limited number—or even a single—noisy image.

{\bf Paper organization.}  Section \ref{sec:related-works} reviews the related work in the literature. Section \ref{sec:lossy-neural-denoising}  gives an overview of the idea of compression-based denoising and then presents our proposed neural compression-based denoiser.  Section \ref{sec:experiments} presents our experimental results. Finally, Section \ref{sec:conclude} concludes the paper.

\section{Related Works}\label{sec:related-works}

Compression-based denoising is a well-established  theoretically founded approach for denoising \cite{donoho2002kolmogorov, weissman2005empirical}. Given the extensive development of   lossy compression algorithms, various compression-based denoising algorithms have been proposed in the literature \cite{saito1994simultaneous,natarajan1995filtering,chang1997image,chang2000adaptive}. For instance, \cite{saito1994simultaneous} employs the Minimum Description Length principle to select the best wavelet coefficients that balance the rate and distortion for denoising the image. Similarly, Natarajan~\cite{natarajan1995filtering} proposes the Occam filter which provides lossy compression of  the source, with the distortion level heuristically selected proportional to the noise level. Chang et al.~\cite{chang1997image,chang2000adaptive} employ optimal soft-thresholding of wavelet coefficients, where the optimized threshold value is claimed to be analogous to the zero-zone in quantizer.  However, while compression-based denoising algorithms have shown to achieve near-optimal performance in denoising some finite, one-dimensional stationary sources \cite{jalali2012denoising}, their performance in image denoising remains sub-optimal compared to other denoising methods.

Data compression, particularly image compression, is a mature research area with over 60 years of history. Recently, the success of deep learning in various machine learning tasks has spurred interest in developing DL-based compression methods. Neural compression networks, introduced as a learning-based approach, have demonstrated superior rate-distortion performance in image compression \cite{
% balle2020nonlinear, 
balle2017endtoend,balle2018variational,minnen2018joint,yang2023introduction
% ,zhu2022transformer, liu2023learned,theis2021on
}. Moreover, in recent years there has also been progress in theoretically understanding the capacity and limitations of neural compression codes. For instance, Wagner et al. \cite{wagner2021neural} show that a neural compression network can achieve the fundamental rate-distortion limits for a simple stochastic source with a one-dimensional latent space. However, Bhadane et al. \cite{bhadane2022neural} find that this optimal performance does not easily extend to sources with a two-dimensional manifold structure. Ozyilkan et al. \cite{ozyilkan2024breaking} highlight the challenges neural networks face in achieving optimal compression when trained on sources with inherent discontinuities. To address this, they propose enhancing neural compression networks by providing Fourier features as input, a technique similar to those used in learning implicit neural representations \cite{sitzmann2020implicit}. 

In this paper, we aim to develop an image denoising algorithm using neural compression networks. A related but fundamentally distinct problem is using these networks to compress noisy images more efficiently, which has a different objective than ours. Compressing noisy images typically requires a higher bit rate, especially when using neural compression models trained on clean images. Therefore, to compress noisy images more efficiently, \cite{testolina2021towards} propose adding convolutional layers in the decoder, inspired by DnCNN \cite{zhang2017beyond}, enabling simultaneous compression and denoising while reducing computational costs at the expense of rate-distortion trade-offs. \cite{larigauderie2022combining} improve upon this by introducing noise level estimation in the latent space, utilizing a two-branch network trained with a distance penalty to allocate fewer bits for noisy images. \cite{xie2024joint} extend this further by incorporating contrastive loss to better align noisy and clean latent codes. Similarly, \cite{ranjbar2022joint} partition the latent code to facilitate separate noisy and clean reconstructions, while \cite{brummer2023importance} investigate training neural compression models directly on noisy images to enhance rate-distortion performance compared to models trained solely on clean data.

\section{Lossy Neural Codes for Denoising}\label{sec:lossy-neural-denoising}

\subsection{Compression-based denoising}

Denoising algorithms leverage  the source structure to differentiate between noise and signal. Image denoising algorithms leverage well-known image structures, such as sparsity in wavelet domain or smoothness,  to achieve this goal. Similarly, data compression algorithms employ the same structures to provide efficient  image representation. Given that at their core, both methods utilize the source structure, one might ask: Given an efficient compression algorithm, can we design an efficient denoising algorithm, that instead of directly dealing with the source structure, employs the compression code to enforce the source model?

\begin{figure}
    \centering
    \includegraphics[width=\linewidth]{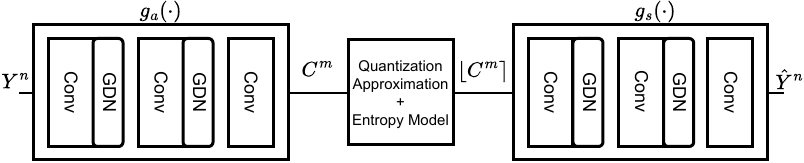}
    \vspace{-0.2in}
    \caption{Structure of the neural compression network used for denoising. \textit{Conv} represents convolutional layer, and \textit{GDN} is the activation function. More details on the structure of network can be found in Section \ref{sec:exp:settings}.}
    \label{fig:neural-net}
\end{figure}

\begin{figure*}
    \centering
    \includegraphics[width=\linewidth]{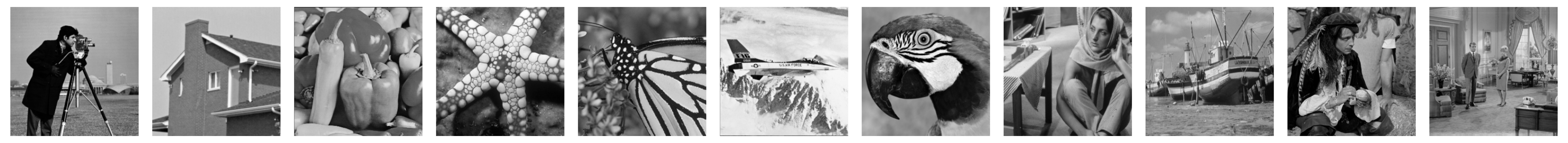}
    \vspace{-0.2in}
    \caption{Eleven test images, with resolutions of $256 \times 256$ for images 1-7 and $512 \times 512$ for image 8-11.}
    \label{fig:set11-test-images}
\end{figure*}
%\textcolor{teal}{Samples of the posterior are actually the structure-enforcer of the previous paragraph, right? Is any sample from posterior desirable then? Don't we need to elaborate a bit more on the connection of previous paragraph to this?} \textcolor{BrickRed}{I agree, I think we could connect the optimal denoiser with compressor from the perspective of joint probability or posterior mean.}

The answer to this question has been shown to be affirmative. Donoho \cite{donoho2002kolmogorov} introduced the concept of minimum Kolmogorov complexity estimator connecting optimal lossy compression and denoising. This result was later refined in \cite{weissman2005empirical} using rate-distortion codes. The authors showed that compressing a noisy process with a family of compression codes that achieves the rate-distortion curve of the noisy process, operating at a distortion level matching the noise level, the joint empirical distribution of the noisy signal and its compressed version converges to the joint distribution of the noisy signal and the source. This result suggests that with additional post-processing, one can achieve  optimal denoising performance \cite{weissman2005empirical} using compression codes.

Despite a solid theoretical foundation, success of compression-based methods in denoising high-dimensional real-valued signals such as images, has been limited. Inspired by neural compression codes and their ability to learn the source distribution from training samples, in the next section we propose a denoising algorithm based on neural compression.

\begin{table*}[t] 
% \small
% \tiny
% \scriptsize
\caption{Denoising performance comparison (PSNR) of the test images in Figure \ref{fig:set11-test-images} under additive Gaussian noise $\mathcal{N}(0, \sigma^2 I)$. \\ DeCompress (single) denotes training our model on a single noisy image for denoising, DeCompress (BSD400) denotes training our model on noisy version of images in the BSD400 dataset.}
\begin{center}
% \begin{sc}
\resizebox{\textwidth}{!}{
\begin{tabular}{cccccccccccccccc}
    \toprule
    &&\multicolumn{7}{c}{$256 \times 256$} &\multicolumn{4}{c}{$512 \times 512$}\\
    \cmidrule(lr){3-9}
    \cmidrule(lr){10-13}
    $\sigma$ & Method & C.man & House & Peppers & Starfish & Monarch & Airplane & Parrot & Barbara & Boats & Pirate & Couple & Average \\
    \midrule 
\multirow{6}{*}{15}
    & BM3D & 31.91 & 34.93 & 32.69 & 31.14 & 31.85 & 31.07 & 31.37 & 33.10 & 32.13 & 31.92 & 32.10 & 32.20 \\ 
    & JPEG2K & 27.14 & 29.22 & 27.86 & 26.63 & 26.77 & 26.61 & 27.14 & 26.78 & 27.97 & 27.78 & 27.44 & 27.39 \\ 
    & Deep Decoder & 26.85 & 30.96 & 25.79 & 28.38 & 28.47 & 25.33 & 27.40 & 24.31 & 28.27 & 28.87 & 27.79 & 27.49 \\ 
    & Noise2Void & 25.68 & 30.05 & 26.60 & 26.41 & 27.34 & 24.38 & 26.41 & 24.36 & 28.73 & 27.33 & 28.36 & 26.88 \\ 
    & \textbf{DeCompress (single)} & 28.68 & 32.81 & 30.89 & 29.41 & 30.01 & 28.24 & 28.16 & 25.94 & 30.94 & 30.74 & 30.56 & 29.67 \\
    & \textbf{DeCompress (BSD400)} & 29.09 & 33.56 & 31.09 & 30.42 & 31.16 & 29.00 & 28.97 & 25.75 & 30.94 & 31.00 & 30.63 & 30.15 \\ 
    \midrule 
\multirow{6}{*}{25}
    & BM3D & 29.45 & 32.86 & 30.16 & 28.56 & 29.09 & 28.42 & 28.93 & 30.71 & 29.90 & 29.61 & 29.71 & 29.76 \\
    & JPEG2K & 24.51 & 27.30 & 24.85 & 24.20 & 24.04 & 23.88 & 24.58 & 24.08 & 25.56 & 25.74 & 25.28 & 24.91 \\ 
    & Deep Decoder & 25.06 & 28.77 & 24.39 & 26.16 & 26.22 & 23.84 & 25.44 & 23.57 & 26.46 & 27.10 & 26.08 & 25.74 \\ 
    & Noise2Void & 23.19 & 27.86 & 24.48 & 23.97 & 25.22 & 22.60 & 23.86 & 22.87 & 26.62 & 25.61 & 26.19 & 24.77 \\ 
    & \textbf{DeCompress (single)} & 26.56 & 30.28 & 27.90 & 26.36 & 27.30 & 26.00 & 26.20 & 24.65 & 28.33 & 28.22 & 27.81 & 27.24 \\
    & \textbf{DeCompress (BSD400)} & 26.76 & 31.09 & 28.23 & 27.20 & 28.13 & 26.64 & 26.72 & 24.38 & 28.38 & 28.50 & 28.01 & 27.64 \\ 
    \midrule 
\multirow{6}{*}{50}
    & BM3D & 26.13 & 29.69 & 26.68 & 25.04 & 25.82 & 25.10 & 25.90 & 27.22 & 26.78 & 26.81 & 26.46 & 26.51 \\
    & JPEG2K & 21.44 & 24.38 & 21.71 & 21.38 & 20.80 & 21.16 & 21.22 & 21.73 & 22.87 & 23.34 & 22.70 & 22.07 \\ 
    & Deep Decoder & 22.80 & 25.98 & 22.60 & 23.30 & 23.12 & 21.10 & 22.52 & 22.60 & 24.07 & 24.73 & 23.93 & 23.34 \\ 
    & Noise2Void & 17.99 & 22.69 & 20.20 & 18.72 & 19.40 & 19.20 & 17.64 & 19.57 & 21.48 & 20.45 & 21.31 & 19.88 \\ 
    & \textbf{DeCompress (single)} & 23.95 & 26.88 & 24.56 & 23.30 & 23.91 & 23.23 & 23.16 & 23.12 & 25.52 & 25.78 & 25.08 & 24.41 \\
    & \textbf{DeCompress (BSD400)} & 24.15 & 27.89 & 25.18 & 23.94 & 24.58 & 23.76 & 24.18 & 23.18 & 25.67 & 26.07 & 25.34 & 24.90 \\ 
    \bottomrule
\end{tabular}
}  % resizebox
% \end{sc}
\label{tab:main_results}
\end{center}
\end{table*}

\subsection{Neural Compression as Denoiser}
Neural image compression algorithms employ an end-to-end trainable neural network to optimize a set of parameters for both the analysis and synthesis nonlinear transforms. Figure \ref{fig:neural-net} shows the proposed neural compression-based denoiser, which receives the noisy signal $Y^n$ as input and generates the denoised signal $\Yh^n$ as output. In this model, $g_a$ and $g_s$ represent the analysis and synthesis transforms, respectively. To train this model, one only requires access to noisy images. The distinction between this model and typical denoising neural nets, such as DnCNN \cite{zhang2017beyond}, is that during training the goal is not to recover  the  clean image. Instead, here, at the training stage, the loss is measured between the input noisy and its reconstruction $\Yh^n$. However, unlike other learning-based methods that do not have access to the clean and noisy image pairs \cite{krull2019noise2void,lehtinen2018noise2noise,kim2021noise2score}, here, due to the regularization done by the compression code, the network does not overfit to  learn an identity map. 

In the neural compression network shown in Figure \ref{fig:neural-net}, the entropy model is a nonparametric density estimation model, which is trained to fit the distribution of the latent code representation of the image \cite{balle2018variational}. 
Let $\lfloor C^m\rceil$ denote the quantized latent code shown in Figure \ref{fig:neural-net}. The bottleneck's rate is defined as
\begin{align*}
    R=\E\left[-\log_2\P(\lfloor C^m\rceil)\right],
\end{align*}
where $C^m$ represents the output of the analysis transform, \ie $C^m=g_a(Y^n)$ and the expectation is computed with respect to the input source $Y^n$. The quantized representation is then mapped to the image domain using a synthesis transform as $\hat{Y}^n=g_s(\lfloor C^m\rceil)$. The rate distortion loss function is then employed as
\begin{align}\label{eq:loss}
    \E[\frac{1}{n}\lVert\hat{Y}^n-Y^n\rVert_2^2]+\lambda R,
\end{align}
with the Lagrangian multiplier $\lambda$ determining the point on the achievable rate-distortion region. 
% \textcolor{blue}{How about moving the next part to the section on network details? }

\section{Experiments}\label{sec:experiments}

We compare our method for denoising against the representative denoiser baselines: 1) BM3D\cite{dabov2007image}, block-matching based on non-local similarity and 3D filtering; 2) Deep Decoder\cite{heckel2018deep}, denoising by regularization from the network's structure; 3) JPEG2K wavelet thresholding\cite{chang1997image, chang2000adaptive
% , dar2018restoration
}, employing JPEG-2000 lossy compression algorithm to denoise via wavelet thresholding; 4) Noise2Void \cite{krull2019noise2void}, a learning based method which only requires access to a dataset of noisy images. The common point among the learning-based baselines is that no pairs of noisy and clean data are needed in the learning phase of the denoisers. The first three approaches are zero-shot, \ie only the noisy image to be denoised is required, while the last baseline (Noise2Void) requires a large training dataset of noisy images. To explore the performance of the proposed neural compression-based denoising, which we name it as \emph{DeCompress}, we study two training settings. Specifically, we train two networks, using 1) a set of noisy images and 2)  a single noisy image. We compare the denoising performance on the clean images shown in Figure \ref{fig:set11-test-images} under different levels of additive noise. 

\begin{figure*}
    \centering
    \includegraphics[width=\linewidth]{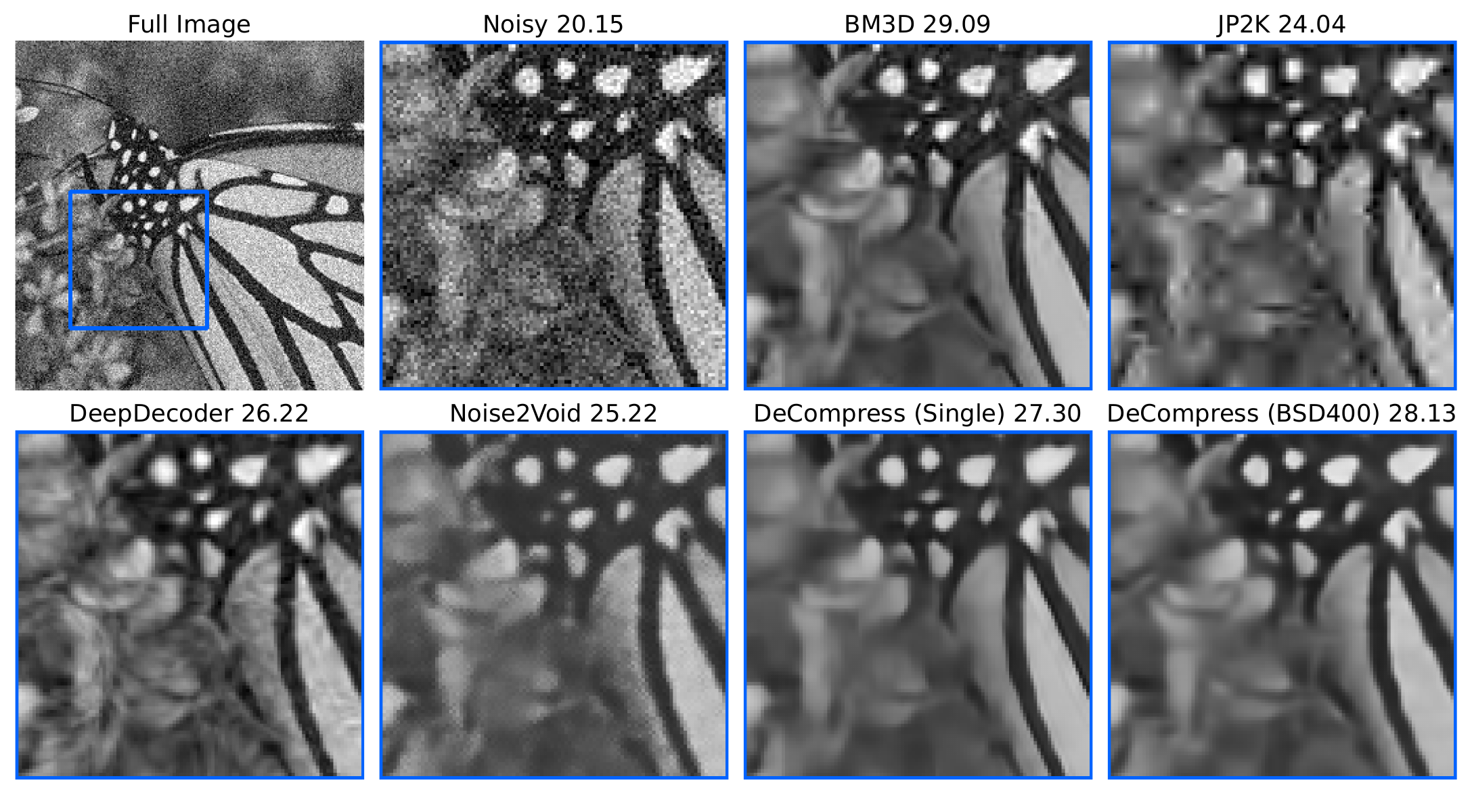}
    \vspace{-0.3in}
    \caption{Visual comparison of denoising methods in Table \ref{tab:main_results} for Monarch image at noise level $\sigma=25$. The PSNR (dB) is above each denoised image.}
    \label{fig:visual-comparison}
\end{figure*}

We numerically compared our two denoisers with others in Table \ref{tab:main_results} using PSNR between the denoised image and its ground-truth, and visually in Figure \ref{fig:visual-comparison}. There are several interesting findings: 1) When the noise level is set correctly in BM3D, the denoising performance is very good that can be used as a good reference for evaluating learning-based denoisers. 2) JPEG2K can be considered as an effective baseline of wavelet thresholding for denoising. 3) Noise2Void's denoising performance is not satisfactory even when trained on a set of noisy images. 4) Our method, neural compression code for denoising, does not rely considerably on the  size of training dataset, and even a single noisy image is sufficient for the networks to learn a compression code with competitive denoising performance.
In the following, we further study different aspects of the neural compression-based denoising. 

\subsection{Training on one vs. many noisy images}
As shown in Table \ref{tab:main_results}, increasing the training data size from a single image to 400 images does not significantly enhance the performance of our compression-based denoisers. This observation aligns with previous studies on neural compression networks \cite{minnen2020channel, zhu2022transformer}. For instance, unlike the generative models, Minnen et al. \cite{minnen2020channel} observed that reconstructed images of random samples of a trained entropy model exhibit significant local structure. Likewise, Zhu et al. \cite{zhu2022transformer} found that, in contrast to the typical behavior of vision transformers, which capture global representations, their transformer-based neural compression models reconstruct each pixel based on only a small region of the input image. They attribute this phenomenon to the rate term, which regularizes the network to focus on local patches during reconstruction. These findings support the idea that the denoising performance of our networks remains largely unaffected by the size of the training dataset and that the training process is unlikely to overfit to a single image.

\subsection{Relationship between $\lambda$ and  $\sigma$?}
We use coefficient $\lambda$ to control the entropy rate of the neural compression code, which directly determines the distortion level as well. As discussed in Section \ref{sec:lossy-neural-denoising}, the rate-distortion operating point of the compression code should be selected as a function of the noise level. For instance, the author in \cite{natarajan1995filtering} states that ``\emph{when the loss level is equal to the true underlying noise level, optimal lossy compression of random process data must exhibit reconstructions which look like samples from the correct Bayesian posterior}.'' Consistently, in our implementations, we use larger $\lambda$ for higher additive noise level. However, the selection of hyperparameter $\lambda$ relies on a good estimation of the noise level, and the networks with different $\lambda$ need to be trained separately for different noise levels. Adaptively selecting $\lambda$ as a function of the noisy image and exploring the possibility of efficiently training compression networks that generalize across different noise levels or the usage of variable-rate neural compression networks \cite{choi2019variable,balle2020nonlinear}, is also an interesting question for future research.

\begin{figure}
    \centering
    \includegraphics[width=0.9\linewidth]{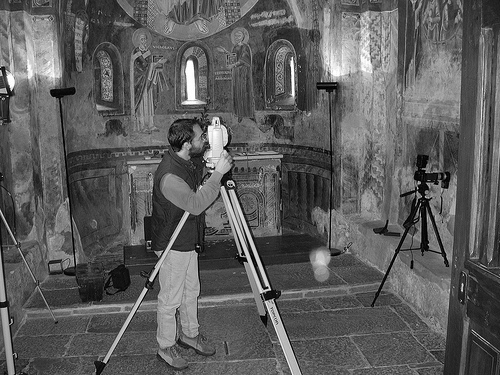}
    \caption{The image randomly selected from ImageNet-1k used for training our neural compression (\textit{ILSVRC2012\_val\_00000059.jpeg}, $500\times375$ pixels).}
    \label{fig:imagenet-image}
\end{figure}

\subsection{Experimental settings}\label{sec:exp:settings}
\paragraph{Neural compression denoisers} We use the network architecture by Ball\'e et al. \cite{balle2018variational} with the Factorized Prior as the entropy model. Our network as shown in Figure \ref{fig:neural-net} consists of three convolutional layers for both the analysis and synthesis networks, each downsizing and upsizing the spatial dimensions by a two at each layer, respectively. The first two convolutional layers in analysis have 256 channels while the last one has 16. All the convolutional layers use a kernel size of $3$ with stride $2$. Lagrangian multiplier of the rate term in loss function (Equation \ref{eq:loss}) is set to $\lambda=300, 1000, 3000$ for noise levels $\sigma=15, 25, 50$, respectively. The activation function applied to the convolutional layer outputs is Generalized Divisive Normalization (GDN) \cite{balle2017endtoend}. Additionally, to avoid zero gradient caused by quantization of the bottleneck during updating parameters of both the analysis network and the entropy model, we use the similar approach proposed by Ball\'e et al. \cite{balle2017endtoend} to approximate the quantization, \ie adding uniform distributed noise $u\in[-1/2,1/2]$ to the bottleneck $C^m$ during training, and rounding to the nearest integer at evaluation. We use Adam optimizer with learning rate of $2\times10^{-4}$ utilizing the neural compression networks implementations provided by the \texttt{CompressAI}\cite{begaint2020compressai} package.

\paragraph{Data preparation} We choose two different datasets to train the neural compression network. First, we use overlapping patches extracted from the image shown in Figure \ref{fig:imagenet-image}. Second, we use the common dataset in learning-based denoising algorithms, BSD400 \cite{martin2001database} consisting of 400 grayscale images of size $180\times180$. In all our experiments with neural compression, we set the input patch size to $16\times16$ with stride step to be 1 both in training and evaluation. During the evaluation, we denoise overlapping patches of size $16\times16$ of the test image, and then take their average with respect to their location to find the overall denoised image.

\paragraph{Baseline settings} The setups for the other denoising algorithms are as follows: For BM3D, the noise level parameter is set equal to the noise level, \ie $\sigma$, in the test noisy image. For Deep Decoder, we train the network for 5000 iterations and calculate the PSNR values with respect to the clean image, reporting the best achieved PSNR in Table \ref{tab:main_results}. For JPEG2K wavelet thresholding \cite{chang1997image}, we use the JPEG-2000 implementation from \texttt{OpenCV}.
% \cite{opencv_library}
The values reported in Table \ref{tab:main_results} are obtained by sweeping over the rate parameter of the JPEG-2000 compression algorithm and selecting the best PSNR achieved with respect to the clean image. For Noise2Void, we use the official implementation provided in the \texttt{CAREamics}\footnote{https://careamics.github.io/0.1/} package. To ensure a fair comparison, we train the Noise2Void networks separately for each noise level from scratch on BSD400, applying data augmentation as described in the original paper \cite{krull2019noise2void}.

\section{Conclusion}\label{sec:conclude}
Inspired by the well-established theoretical connection between optimal lossy compression and denoising, we propose neural compression-based image denoising method. Our experimental results show that a neural network optimized with rate-distortion loss over a dataset of noisy image(s) is capable of delivering competitive denoising performance to  zero-shot learning-based deneoising methods.

Over the past decade, compression-based algorithms  have  been studied, both theoretically and algorithmically, for a  range of  inverse problems beyond denoising  \cite{jalali2016compression,rezagah2017compression,dar2018restoration,beygi2019efficient}. Algorithmically exploring application of neural compression techniques  in solving these inverse problems is an intriguing and promising  direction for future research.

%%%%%%
%% Appendix:
%% If needed a single appendix is created by
%%
%\appendix
%%
%% If several appendices are needed, then the command
%%
% \appendices
%%
%% in combination with further \section commands can be used.
%%%%%%

\section*{Acknowledgment}
A.Z., X.C., S.J. were supported by NSF CCF-2237538.

%%%%%%
%% To balance the columns at the last page of the paper use this
%% command:
%%
%\enlargethispage{-1.2cm} 
%%
%% If the balancing should occur in the middle of the references, use
%% the following trigger:
%%
%\IEEEtriggeratref{7}
%%
%% which triggers a \newpage (i.e., new column) just before the given
%% reference number. Note that you need to adapt this if you modify
%% the paper.  The "triggered" command can be changed if desired:
%%
%\IEEEtriggercmd{\enlargethispage{-20cm}}
%%
%%%%%%

%%%%%%
%% References:
%% We recommend the usage of BibTeX:
%%
\bibliographystyle{IEEEtran}
\bibliography{refs}
%%
%% where we here have assumed the existence of the files
%% definitions.bib and bibliofile.bib.
%% BibTeX documentation can be obtained at:
%% http://www.ctan.org/tex-archive/biblio/bibtex/contrib/doc/
%%%%%%

%% Or you use manual references (pay attention to consistency and the
%% formatting style!):

\end{document}

%% file: defns.tex
\usepackage{xspace}
\usepackage{bbm}
\newcommand{\Nc}{\mathcal{N}}

\newcommand{\Xc}{\mathcal{X}}

\newcommand{\Yh}{{\hat{Y}}}

\let\E\relax
\DeclareMathOperator\E{E}
\let\P\relax
\DeclareMathOperator\P{P}
\newcommand\ie{i.e.,\xspace}
\def\textiid{i.i.d.\@\xspace}
\newcommand\iid{\ifmmode\text{ i.i.d. } \else \textiid \fi}